\documentclass[prd,preprint,superscriptaddress]{revtex4}
\usepackage{amsmath}
\usepackage{amssymb}
\usepackage{revsymb}
\usepackage{graphicx}

\newcommand{\be}{\begin{equation}}
\newcommand{\ee}{\end{equation}}
\newcommand{\ben}{\begin{eqnarray}}
\newcommand{\een}{\end{eqnarray}}
\newcommand{\n}{\label}

\begin{document}

\title{Dark energy and the generalized second law}
\author{Germ\'{a}n Izquierdo\footnote{
E-mail address: german.izquierdo@uab.es}}
\affiliation{Departamento de F\'{\i}sica, Universidad Aut\'{o}noma de Barcelona, 08193
Bellaterra (Barcelona), Spain}
\author{Diego Pav\'{o}n\footnote{
E-mail address: diego.pavon@uab.es}}
\affiliation{Departamento de F\'{\i}sica, Universidad Aut\'{o}noma de Barcelona, 08193
Bellaterra (Barcelona), Spain}

\begin{abstract}
We explore the thermodynamics of dark energy taking into account
the existence of the observer's event horizon in accelerated
universes. Except for the initial stage of Chaplygin gas dominated
expansion, the generalized second law of gravitational
thermodynamics is fulfilled and the temperature of the phantom
fluid results positive. This substantially extends the work of
Pollock and Singh \cite{pollock} on the thermodynamics of
super--inflationary expansion.
\end{abstract}

\maketitle
\section{Introduction}
Nowadays there is an ample consensus on the observational side
that the Universe is undergoing an accelerated expansion likely
driven by some unknown fluid (called dark energy) with the
distinguishing feature of violating the strong energy condition,
$\rho + 3P > 0$ \cite{consensus}. The strength of this
acceleration is presently a matter of debate mainly because it
depends on the theoretical model employed when interpreting the
data. While most model independent analyses  suggest it to be
below the de Sitter value \cite{independent} it is certainly true
that the body of observational data  allows for a wide parameter
space compatible with an acceleration larger than de Sitter's
\cite{caldwell1,steen}. If eventually this proves to be  the case,
the fluid driving the expansion would violate  not only the strong
energy condition but the dominant energy condition, $\rho + P >%
0$, as well. In general, fluids of such characteristics, dubbed
``phantom" fluids, face some theoretical difficulties on their own
such as quantum instabilities \cite{carroll}. Nevertheless,
phantom models have attracted much interest, among other things
because in their barest formulation they predict the Universe to
end from an infinite expansion in a finite time (``big rip"),
rather than in a big crunch, preceded by the ripping apart of all
bound systems, from galaxy clusters down to atomic
nuclei\footnote{The big rip was first discovered by Davies
\cite{first}, it was later rediscovered and popularized by
Caldwell {\em et al} \cite{caldwell1}.}. While this scenario might
look weird, given our incomplete understanding of the physics
below $\rho + P = 0$ and the scarcity of reliable observational
data, it should not be discarded right away; on the contrary we
believe it warrants some further consideration.

Recently, it has been demonstrated that if the expansion of the
Universe is dominated by phantom fluid, black holes will decrease
their mass and eventually disappear altogether \cite{babichev}. At
first sight this means a threat for the second law of
thermodynamics as these collapsed  objects are the most entropic
entities of our world. This short consideration spurs us to
explore the thermodynamic consequences of phantom--dominated
Universes. In doing so one must take into account that ever
accelerating universes have a future event horizon (or
cosmological horizon). Since the horizon implies a classically
unsurmountable barrier to our ability to see what lies beyond it,
it is natural to attach an entropy to the horizon size (i.e., to
its area) for it is a measure of our ignorance about what is going
on in the other side. However,  this has been proved in a rigorous
manner  for the Sitter horizon only \cite{gary} which, in
addition, has a temperature proportional to the Hubble expansion
rate, $T_{H} = \hbar H/(2\pi k_{B})$. Nevertheless, following
previous authors \cite{first, pauld,pollock,brustein} here we
conjecture that this is also true for non-stationary event
horizons. While certainly this is a bold assumption that we cannot
justify at present we believe the reasonableness of the results
lend support to it.

Phantom expansions of pole-like type
\\
\be
a(t) \propto \frac{1}{(t_{*} - t)^{n}} \qquad (t \leq t_{*}, \; \;
0 < n = \mbox{constant})\, ,
\n{phexpansions}
\ee
\\
as proposed by Pollock and Singh \cite{pollock} and Caldwell
\cite{caldwell2}, arise when the equation of state parameter $w
\equiv P/\rho = \mbox{constant} < -1$. Current cosmological
observations hint that $w$ may be as lower as $-1.5$ \cite{wdata}.
In what follows it will be useful to bear in mind that $ n = -
2/[3(1+w)] > 0$.

From the above equation for the scale factor it is seen that
the Hubble expansion rate augments
\\
\be
H(t) \equiv \frac{\dot{a}}{a} = \frac{n}{t_{*} - t}\,  ,
\n{hubble1}
\ee
\\
whereby the radius of the observer's event horizon
\\
\be
R_{H} = a(t)\int_{t}^{\infty}{\frac{dt'}{a(t')}} = a(t)\int_{t}^{t_{*}}{\frac{dt'}{a(t')}}
= \frac{t_{*} - t}{1+n} = \frac{n}{1+n} H^{-1}  \qquad (c = 1)\,
\n{event1}
\ee
\\
decreases with time, i.e., $\dot{R}_{H}< 0$, and vanishes altogether at the big rip
time $t_{*}$.

Consequently the horizon entropy,
\\
\be
S_{H} = \frac{k_{B}}{4} \, \frac{{\cal A}}{\ell_{pl}^{2}}\, ,
\n{hentropy}
\ee
\\
where ${\cal A} = 4\pi R_{H}^2$ denotes the area of the horizon and
$\ell_{pl} \equiv \sqrt{\hbar G/c^{3}}$ is the Planck's length,
diminishes with time, $\dot{S}_{H} < 0$. This is only natural since
for spatially flat FRW phantom-dominated universes one has
\\
\[
\dot{H} = -\frac{1}{2} \, \kappa \, (\rho+ P) > 0 \qquad \quad
(\kappa = 8\pi\, G)\, ,
\]
\\
as these fluids violate the dominant energy condition. Then the
question arises, ``will the generalized second law (GSL) of
gravitational thermodynamics, $\dot{S}\, + \, \dot{S}_{H} \geq 0$,
be satisfied?" The target of this work is to answer this question.
According to the GSL the entropy of matter and fields inside the
horizon plus the entropy of the event horizon cannot decrease with
time. To our knowledge the first authors who considered this
question assuming energy sources that violate the dominant energy
condition were Pollock and Singh \cite{pollock}. They studied the
super-inflation models of Starobinsky \cite{starobinsky} and Shafi
and Wetterich \cite{shafi} in the regime that they departed only
slightly from de Sitter expansion and found that while the former
fulfills the GSL the latter does not. We extend the work of these
authors by considering other models of accelerated expansion
(phantom and non-pantom) not restricting ourselves to small
deviations from de Sitter. As it turns out, except for the
Chaplygin gas dominated universe (section III) the GSL is
fulfilled in all the instances explored.

Since sooner or later the expansion will get dominated by the dark
energy fluid we will neglect, for the sake of transparency and
simplicity, all other sources of energy (e.g., matter, radiation,
and so on). For the interplay between ordinary matter (and/or
radiation) and the cosmological event horizon, see Refs.
\cite{first,pauld,brustein,relic}.

\section{Dark energy with constant $w$ \label{phcntw}}
We begin by considering a phantom fluid inside the cosmological
event horizon of a comoving observer. Its entropy can be related
to its energy and pressure in the horizon by Gibbs' equation
\\
\be T \, dS = dE\, + \, P \, d \left(\textstyle{4\over{3}}\pi
R_{H}^{3}\right)\, . \n{gibbs1} \ee
\\
In this expression, owing to the fact that the number of ``phantom
particles" inside the horizon is not conserved, we have set the
chemical potential to zero. From the relation $E = (4\pi/3) \rho
R_{H}^{3}$, together with the Friedmann equation $\rho =
(3/\kappa) H^{2}$, Eq. (\ref{event1}), and the equation of state
$P= w \rho$  (with $w < -1$), we get
\\
\be T \, dS = -\frac{8\pi}{\kappa}\, \frac{n}{1+n}\, dR_{H}.
\n{gibbs2}
\ee
\\
Since $dR_{H} <0$ the phantom entropy increases with expansion (so
long as $T >0$).

One may argue that it is wrong to apply Gibbs equation to a fluid
that in reality is a phenomenological representation of a scalar
field assumed to be in a specific state, and that in such a case
its  entropy should strictly vanish. This reasoning could be
correct if we were supposing that the fluid phenomenologically
represents just one single field and this one in a pure state, but
this is far from being the more natural assumption. We see the
fluid as a phenomenological representation of a mixture of fields,
each of which may or may not be in a pure state but the overall
(or effective) ``field" is certainly in a mixed state and
therefore entitled to an entropy. This is the case, for instance,
of assisted inflation \cite{assisted}.

To proceed further, we must specify the temperature of the phantom
fluid. The only temperature scale we have at our disposal is the
temperature of the event horizon, which we assume to be given by
its de Sitter expression \cite{gary}
\\
\be T_{H} = \frac{\hbar}{2 \pi \, k_{B}}\, \frac{n}{1+n}
\frac{1}{R_{H}} \, , \label{TH1} \ee
\\
though in our case $\dot{H} > 0$. Thus, it is natural to suppose
that $T \propto T_{H}$ and then figure out the proportionality
constant. As we shall see below, this choice is backed by the
realization that it is in keeping with the holographic principle
\cite{holographic}. As in Ref.\cite{pollock}, the simplest choice
is to take the proportionality constant as unity which means
thermal equilibrium with the event horizon, $T = T_{H}$. In
general, two systems must interact for some length of time before
they can attain thermal equilibrium. In the case at hand, the
interaction certainly exists as any variation in the energy
density and/or pressure of the fluid will automatically induce a
modification of the horizon radius via Einstein's equations.
Moreover if $T \neq T_{H}$, then energy would spontaneously flow
between the horizon and the fluid (or viceversa), something at
variance with the FRW geometry.

After integrating Eq. (\ref{gibbs2}) and bearing in mind that $S%
\rightarrow 0$ as $t \rightarrow t_{*}$, the entropy of the
phantom fluid can be written as $S = - k_{B}\,{\cal A}/(4\,
\ell_{pl}^{2})$.

Some consequences follow: $(i)$ The phantom entropy is a negative
quantity, something already noted by other authors
\cite{pollock,snegative}, and equals to minus the entropy of a
black hole of radius $R_{H}$. $(ii)$ It bears no explicit
dependence on $w$ and, but for the sign, it exactly coincides with
the entropy of the cosmological event horizon. $(iii)$ Since  $ -S
\propto E^{2}$ it is not an extensive quantity. Note that $S$ does
not scale with the volume of the horizon but with its area. One
may argue that the  first consequence was to be expected since
(for $T> 0$) it readily follows from  Euler's equation $T \, s =
\rho +P$, $s$ being the entropy density. However, it is very
doubtful that Euler's equation holds for phantom fluids since it
is based on the extensive character of the  entropy of the system
under consideration \cite{callen} and we have just argued that for
phantom fluids this is not the case.

Two further consequences are as follows: $(iv)$ The sum $S+ S_{H}$
vanishes at any time,  therefore the GSL is not violated -the
increase of the (negative) phantom entropy exactly offsets the
entropy decline of the event horizon. $(v)$ $\mid S \mid$
saturates the bound imposed by the holographic principle
\cite{holographic}. The latter asserts that the total entropy of a
system enclosed by a spherical surface cannot be larger than a
quarter of the area of that surface measured in Planck units. (For
papers dealing with the holographic principle in relation with
dark energy, see \cite{cohen} and references therein). In this
connection, it is interesting to see that if the equation of state
were such that the entropy obeyed $S^{*}= \varkappa \,  S$, with
$\varkappa $ a positive--definite constant, then the GSL would
impose $1 \leq \varkappa$. This leads us to conjecture  that the
entropy of phantom energy is not bounded from above but from
below, being $- S_{H}$ its lower limit.

Nevertheless, in a less idealized cosmology one should consider
the presence of other forms of energy, in particular of black
holes and the decrease in entropy of these objects
\cite{babichev}. It is unclear whether in such scenario the GSL
would still hold its ground. Nonetheless, one may take the view
that the GSL may impose an upper bound to the entropy stored in
the form of black holes. At any rate, the calculation would be
much more involved and lies beyond the scope of this paper. Among
other things, one should take into account that the scale factor
would not obey such a simple expansion law as (\ref{phexpansions})
and that the black holes would be evaporating via Hawking
radiation which would also modify the expansion rate and,
accordingly, the horizon size.

A straightforward and parallel study for a non--phantom dark
energy-dominated universe with constant parameter of state (lying
in the range $-1 < w <-1/3$), shows that in this case
$\dot{S}+\dot{S}_{H} = 0$ as well\footnote {There exists in the
literature a host of dark energy models with $w =$ constant $%
> -1$, see e.g. \cite{wconstant} and references therein. Any two
of them with the same $w$ lead to identical result with regard to
the GSL.}. There are two main differences, however; on the one
hand the entropy of the fluid decreases while the area of event
horizon augments, and on the other hand $S \geq 0$. The latter
comes from adopting the view that the dark energy must vanish for
$T \rightarrow 0$ (Planck's statement of the third law of
thermodynamics) and realizing that this happens for $t \rightarrow
\infty$. Again, $S$ saturates the bound imposed by the holographic
principle.

Obviously, one might adopt the view that the phantom temperature
should be negative \cite{pedro}. But, as mentioned above, this
would destroy the FRW geometry. On the other hand, negative
temperatures are linked to condensed matter systems whose energy
spectrum is bounded from above and may therefore exhibit the
phenomenon of population inversion, i.e., their upper energy
states can be found more populated than their lower energy states
\cite{populated}. However, all models of phantom energy proposed
so far assume some or other type of scalar field with no upper
bound on their energy spectrum. In addition, while population
inversion is a rather transient phenomenon the phantom regime is
supposed to last for cosmological times. Moreover, bearing in mind
that $w< -1$ and $dR_{H} < 0$, it follows from Eq. (\ref{gibbs2})
that if $T$ were negative, then the phantom entropy would decrease
with expansion and the GSL would  be violated.

\section{Dark energy with variable $w$}
One may argue that the result $\dot{S}+\dot{S}_{H} = 0$ critically
depends on the particular choice of the equation of state
parameter $w$ to the point that if  it were not constant, then the
GSL could be violated. Here we shall explore this issue taking the
phantom model with $w$ dependent on time of Sami and Toporensky
\cite{sami}.

In this model a scalar field, $\phi(t)$, with negative kinetic
energy, minimally coupled to gravity sources with energy density
and pressure given by%
\begin{equation*}
\rho =-\frac{\overset{.}{\phi}^{2}(t)}{2}+V(\phi ),\text{\qquad }%
P =-\frac{\overset{.}{\phi }^{2}(t)}{2}-V(\phi ),
\end{equation*}%
respectively, is adopted.
The equation of state parameter
\begin{equation*}
w=\frac{P}{\rho}=\frac{\frac{\overset{.}{\phi }^{2}(t)}{2}%
+V(\phi )}{\frac{\overset{.}{\phi }^{2}(t)}{2}-V(\phi )},
\end{equation*}%
is lower than $-1$ as long as $\rho$ is positive and depends on
time.

\begin{figure}[tbp]
\includegraphics*[scale=0.5,angle=0]{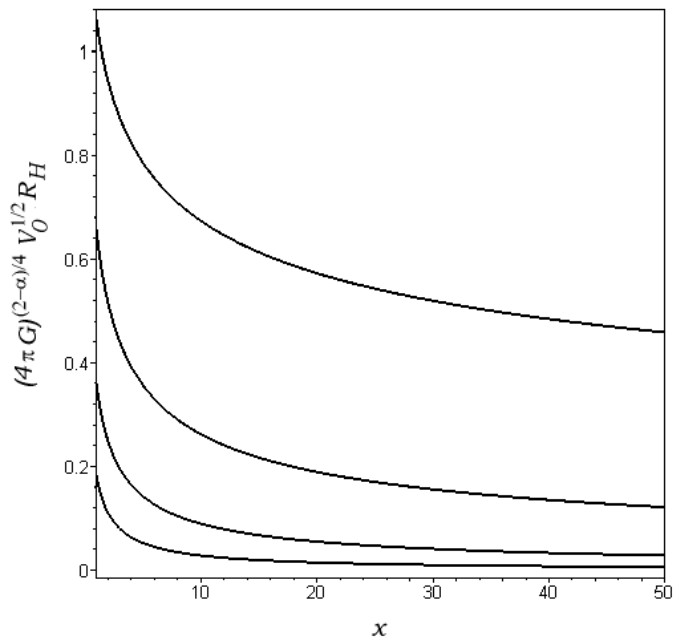}
\caption{Radius of the event horizon {\it vs} $x$ for $\alpha= $
$1$, $2$, $3$ and $4$, from top to bottom. In all the cases $R_{H}
\rightarrow 0$ as $t \rightarrow \infty$. For convenience sake we
have set $x_{i}=1$.} \label{horvsa}
\end{figure}

When the kinetic energy term is subdominant (``slow climb"
\cite{sami}) the evolution equation of the phantom field
simplifies to
\\
\begin{equation}
3H\overset{.}{\phi }(t)\simeq \frac{dV(\phi )}{d\phi }.%
\text{\qquad }
\label{fielev1}
\end{equation}%
\\
Assuming also the power law potential $V(\phi) = V_{0}\,
\phi^{\alpha}$ with $\alpha $ a constant, restricted to the
interval $0<\alpha \leq 4$ to evade the big rip, the equation of
state parameter reduces to $w \simeq - 1 - \frac{\alpha}{6x}$,
where $x = \kappa\, \phi^{2}/(2 \alpha)$ is a dimensionless
variable, and the field is an ever increasing function of time,
namely,
\\
\ben
\phi(t)&=& \left[ \phi _{i}^{\frac{4-\alpha }{2}}+\sqrt{\frac{V_{0}}{3 \kappa}}%
\, \frac{\left( 4-\alpha \right) \alpha }{2}\left( t-t_{i}\right) \right] ^{%
\frac{2}{4-\alpha }} \qquad \quad (0 < \alpha <4), \\
\phi(t)&=& \phi_{i} \, \exp\left[4 \sqrt{\frac{V_{0}}{3\kappa}}\,
(t-t_{i})\right] \qquad (\alpha = 4), \label{field} \een
\\
where the subscript $i$ stands for the time at which the phantom energy
begins to dominate the expansion. Thus $w$ increases with expansion
up to the asymptotic value $-1$. Likewise the scale factor obeys
\\
\be
a(\phi) = a_{i}\, \exp \left[ \frac{\kappa}{2 \alpha}\, (\phi^{2} - \phi_{i}^{2}) \right]\, .
\label{scalefactor}
\ee
As it should, the Hubble factor
\\
\be
H = \sqrt{\frac{2 V_{0}}{3}} \, \left(\frac{\kappa}{2}\right)^{\frac{1}{2}-\frac{\alpha}{4}}
\alpha^{\alpha/4} \, x^{\alpha/4}
\label{hfactor}
\ee
\\
augments with expansion while the cosmological event horizon
\\
\be
R_{H} = x^{\alpha/4}\, \Gamma \left(1- \frac{\alpha}{4}, x\right) \, \frac{e^{x}}{H}
\label{event2}
\ee
\\
decreases monotonically to vanish asymptotically -see Figure
\ref{horvsa}. Here $\Gamma \left( \frac{4-\alpha }{4},x\right) $
is the incomplete Gamma function \cite{Abram}.
\\

As in the preceding section, the entropy of this phantom fluid can
be obtained by using Gibbs's equation and $T = T_{H} = (c \,
\hbar/2\pi k_{B})H$. Again, we get a negative quantity that
increases with time, namely,
\\
\be S = - \frac{8k_{B}\, \pi^{2}}{\hbar \,\kappa}\,  C^{2}\,
\int_{x}^{\infty}{} \Gamma^{2}\left( 1- \frac{\alpha}{4},\,
x\right)\, \frac{\alpha}{2x}\, e^{2x}\, dx \, , \label{sphantom2}
\ee
\\
here $C= \sqrt{3/(2V_{0})}\,
(\kappa/2)^{-\frac{1}{2}+\frac{\alpha}{4}}\, \alpha^{-\alpha/4}$.

From the above expressions, it follows that
\\
\be \dot{S} + \dot{S}_{H} = \frac{8 k_{B}\, \pi^{2}}{\hbar\,
\kappa}\, C \, \left[\Gamma \left( \frac{4-\alpha }{4},\,
x\right)\, e^{x} (2 + \frac{\alpha}{2x})- 2 x^{-\alpha/4}
\right]\, R_{H}\, \dot{x} \, . \label{gsl2} \ee
\\
Since the quantity in square brackets is positive--definite for
any finite $x$ and $\dot{x}>0 $, the GSL in the form $\dot{S} +
\dot{S}_{H} \geq 0 $ is satisfied. The equality sign occurs just
for $t \rightarrow \infty $, i.e., when $R_{H}$ vanishes.

\begin{figure}[tbp]
\includegraphics*[scale=0.6,angle=0]{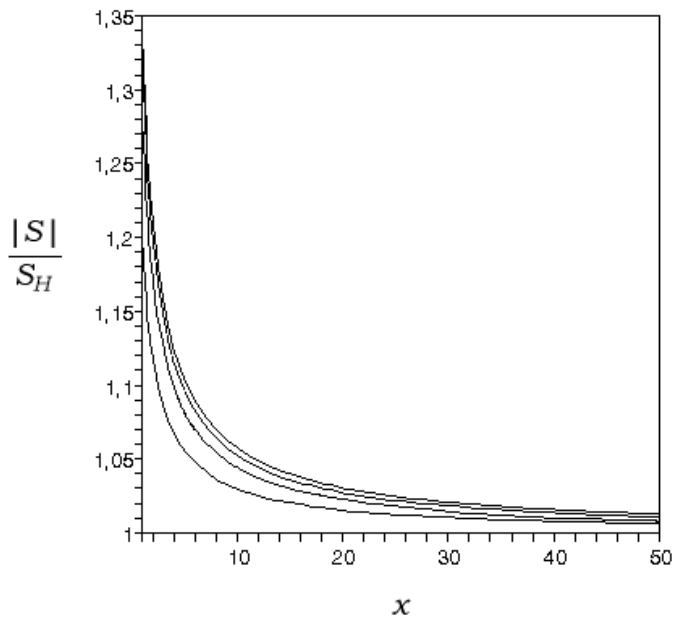}
\caption{Evolution of the ratio between the absolute value of the
phantom entropy and the horizon entropy for the model of Ref.
\cite{sami}. The graphs correspond successively to $\alpha$ values
$1$, $2$, $3$, and $4$ from bottom to top. For convenience sake we
have set $x_{i} = 1$ in drawing the graphs. All the curves
asymptote to $x$ = 1.} \label{sami-holographic}
\end{figure}

As figure \ref{sami-holographic} illustrates we have that $\mid%
S\mid \geq S_{H}$; however, strictly speaking the holographic
principle is respected as $S$ is always lower than $S_{H}$.
\\

As an example of non-phantom dark energy with variable $w$ we
consider the Chaplygin gas \cite{chaplygin}. This fluid is
characterized by the equation of state $P = - A/\rho$. Integration
of the energy conservation equation yields $\rho =
\sqrt{A+(B/a^{6})}$, whereby $w = -A \, a^{6}/(A\, a^{6}\,+B)$.
Here $a$ is the scale factor normalized to its present value and
$A$ and $B$ denote positive--definite constants. Accordingly, this
fluid interpolates between cold matter, for large redshift, and a
cosmological constant, for small redshift. This has led to propose
it as a candidate to unify dark matter and dark energy
\cite{unify}.

\begin{figure}[tbp]
\includegraphics*[scale=0.5,angle=0]{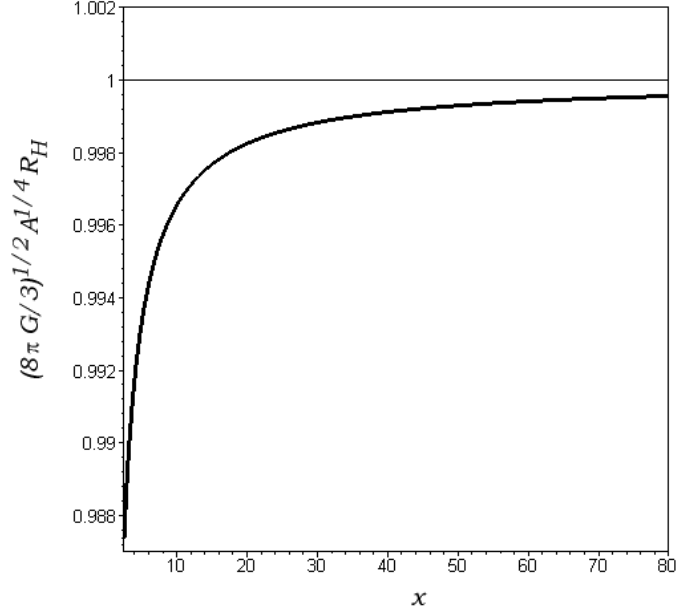}
\caption{Evolution of the adimensionalized radius of the event horizon in
the Chaplygin gas model. It tends to unity for $x \rightarrow \infty$.}
\label{rh-chaplygin}
\end{figure}

The radius of the cosmological event horizon reads
\\
\be
R_{H} = \sqrt{\frac{\kappa}{3}} \, \frac{2}{A^{1/4}}\, x^{1/6} \,
\left[ 1.4712 - x^{1/12} \, \, _{2}F_{1}\left(\frac{1}{12}, \frac{1}{4}, \frac{13}{12}; -x\right) \right]\, ,
\label{event3}
\ee
\\
where $_{2}F_{1}$ is the hypergeometric function and $x = A\, a^{6}/B$ is a
dimensionless variable. As figure \ref{rh-chaplygin} illustrates,
$\mbox{Lim}_{\, x \rightarrow \infty} \,R_{H} = \sqrt{3/\kappa}\, A^{-1/4}$
which is nothing but the value taken by $H^{-1}$ in that limit.
This was to be expected: for large scale factor the  Chaplygin
gas expansion goes over de Sitter's.
\\

The horizon temperature is
\\
\be T_{H} = \frac{\hbar \, c}{2 \pi \, k_{B}}\,
\sqrt{\frac{\kappa}{3}} \, A^{1/4}\,
\left(\frac{1+x}{x}\right)^{1/4}, \label{TH2} \ee
\\
while the entropy of the fluid and the entropy of the horizon evolve as
\\
\be dS= - \frac{48k_{B}\, \pi^{2}}{\hbar \kappa^{2}\, A^{1/2}} \,
\frac{1}{x^{2/3}(1+x)} \left[1.4712 - x^{1/2}\,\,
_{2}F_{1}\left(\frac{1}{12}, \frac{1}{4}, \frac{13}{12}; -x\right)
\right]^{2} \, dx \, , \ee \label{schaplygin}
\\
and
\\
\ben
dS_{H} = \frac{16 k_{B}\, \pi^{2}}{\hbar \kappa^{2}\, A^{1/2}} \, x^{1/6}\,
\left[1.4712 - x^{1/2}\, \,  _{2}F_{1}\left(\frac{1}{12}, \frac{1}{4}, \frac{13}{12}; -x\right) \right] \\
\nonumber
\times \left\{2x^{-5/6} \left[1.4712  - x^{1/2}\, \,  _{2}F_{1}\left(\frac{1}{12},
\frac{1}{4}, \frac{13}{12}; -x\right)\right] - \frac{1}{x^{3/4}\, (1+x)^{1/4}} \right\} \, ,
\label{sh3}
\een
\\
respectively. The two latter equation imply that $\dot{S}\, + \,%
\dot{S}_{H}  \geq 0$ for $x \geq 2.509$ or  equivalently, for
$\rho_{i}\leq 1.18 A^{1/2}$. Thus, there is an early period in
which the GSL seems to fail. However, this might be seen not as a
failure of the GSL but as an upper bound on the initial value of
the energy density of the Chaplygin gas.

Notice that, $S$ is a positive, ever--decreasing function of the
scale factor that complies with the holographic bound $S \leq
S_{H}$.

\section{The issue of negative entropies}
The fact that $S <0$ looks troublesome. In particular, it tells us
that the entropy of the phantom fluid is not to be understood as a
measure of the number of microstates associated to the macroscopic
state of the fluid, i.e., the well-known statistical mechanics
formula $S = k_{B} \, \ln W$ breaks down. However, the fact
remains that the entropy of systems violating that condition ($w%
<-1$) is found to be negative (see also, \cite{pollock,snegative})
whence one is rightly entitled to wonder whether these systems can
be realized in nature. While the answer to this question lies
beyond our present capabilities  we may try to advance some ideas,
pending a deeper study. First, we wish to recall that no physical
law asserts that the entropy of a system has to be positive, the
second law of thermodynamics simply establishes that the entropy
of isolated systems cannot decrease. Secondly, one may draw a
rough formal analogy between dark energy fluids with equation of
state $P = w \rho$ and a very well known physical system, the
mono-atomic ideal gas. The entropy of the latter is given by the
Sackur--Tetrode equation (see Eq. (9.54) of Ref. \cite{huang}),
which can be written as
\\
\[
\frac{\sigma}{k_{B}} = \frac{3}{2} - \ln \left[\left(\frac{2 \pi
\hbar^{2}}{m k_{B}\, T} \right)^{3/2} \, n \right]\, ,
\]
\\
where $\sigma$ is the entropy per particle, $m$ the particle mass,
and $n$ the particle number density. Clearly, by lowering the
temperature below $T_{c} = (2\, \pi \hbar^{2}/e\, m) \, n^{2/3}$
the entropy becomes negative and the pressure, $P = 2 n%
\varepsilon/3$, with $\varepsilon$ the energy per particle,
decreases accordingly (though it stays positive). Thus, by
lowering $P$ (at fixed $n$) negative entropies can be formally
attained. Obviously, one can argue that before a negative entropy
state could be reached quantum effects would invalidate the
approximations leading to the  Sackur-Tetrode equation, and that,
in any case, the gas would condensate ahead of that state.

This is somewhat analogous to a dark energy fluid whose pressure
is steadily decreased by lowering $w$. When the $w = -1$ divide is
crossed, the entropy becomes negative but simultaneously quantum
instabilities might destroy the state \cite{carroll}.

So, we are at a cross-road: either states violating the dominant
energy condition are to be banned (as they entail negative
entropies) or statistical mechanics is to be generalized to
encompass these states. It is for the reader to decide which path
to follow.

\section{Quasi-duality between phantom and non-phantom thermodynamics}
\begin{figure}[tbp]
\includegraphics*[scale=0.5,angle=-90]{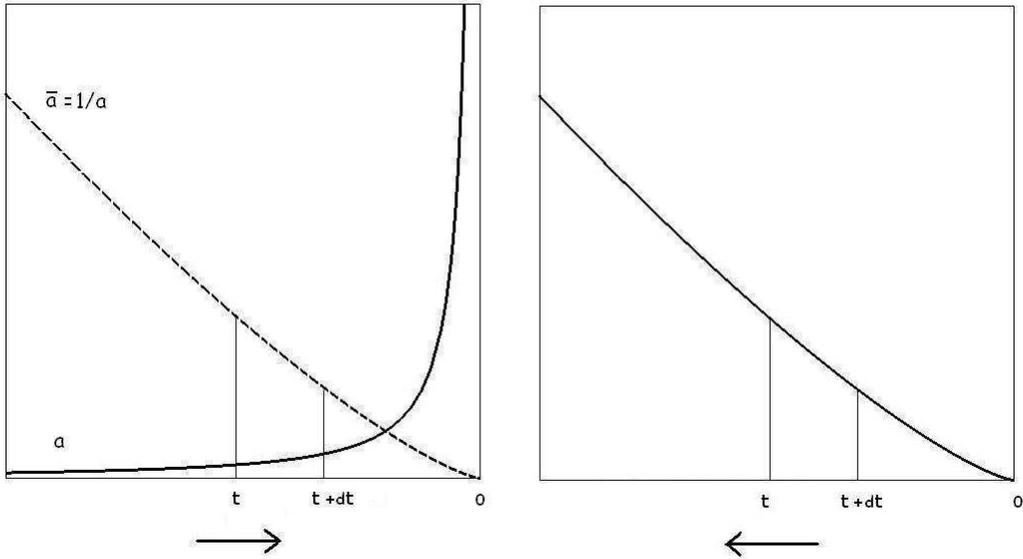}
\caption{The left panel depicts the scale factor of the original
phantom dominated universe (solid line). The dashed line shows the
evolution of the scale factor dual of this phantom DE model, a
contracting universe. In both models (the phantom and its dual)
the time grows from left to right as tends to $t=0^-$ and the
instant $t+dt$ lies at the future of the instant $t$. In the right
panel, the time direction is reversed ($t+dt$ lies at the past of
$t$). Under this operation, the contracting model of the left
panel becomes the non-phantom DE dominated universe of the right
panel if $-5/3<w<-1$.}
\label{duality}
\end{figure}

The duality transformation
\\
\be
\left\{
\begin{array}{ll}
H&\rightarrow\overline{H}=-H\\
\rho+P&\rightarrow \overline{\rho}+\overline{P}=-(\rho+P)
\end{array}
\right.\ee%
leaves the Einstein's equations%
\be
\begin{array}{ll}
H^2=\frac{\kappa}{3}\rho,&\\
\dot{\rho}+3H(\rho+P)=0,&\\
\dot{H}=-\frac{1}{2}\kappa(\rho+P),& \label{einsteinequations}
\end{array}
\ee%
 of spatial FRW universes invariant and the scale factor
transforms as $a\rightarrow\overline{a}=1/a$ \cite{duality}. Thus,
a phantom dominated universe with scale factor $a=1/(-t)^n$ (for
simplicity we assume a phantom scale factor as Eq.
(\ref{phexpansions}) with $t_*=0$ and $t<0$ that tends to $0^-$)
becomes a contracting universe whose scale factor obeys
$\overline{a}=(-t)^n$. This universe begins contracting from an
infinite expansion at the infinite past and collapses to a
vanishing scale factor for $t=0^-$ (left panel of Fig.
\ref{duality}).

Reversing the direction of time (i.e., the future of $t$, $t +
dt$, transforms into $t -dt$), the contracting universe is mapped
into an expanding one whose scale factor grows with no bound as
$t$ tends to $-\infty$. Notice that the first transformation takes
the universe from expanding to contracting and the second
transformation makes the universe expand again, i.e., the final
cosmology has $\widetilde{\overline{H}} > 0$. It should be
remarked that the latter transformation also preserves Eqs.
(\ref{einsteinequations}).

Under these two successive operations, the final and the initial
equation of states parameters are related by $w \rightarrow
\widetilde{\overline{w}}=-(w+2)$. Consequently, phantom dominated
universes with $n> 1$, (i.e., $-5/3 < w < -1$) are mapped into
non-phantom dark energy dominated universes (see Fig.
\ref{duality}). Therefore, both of them, the original and the
transformed universes, have an event horizon and entropies whose
relation we derive below.

The radius of the event horizon of the original universe is $R_H =
(-t)/(1+ n)$ meanwhile the radius of the event horizon of the
transformed universe is $\widetilde{\overline{R}}_H = (-t)/(n - 1)
= R_H[(1 + n)/(1-n)]$. The transformations preserve the horizon
temperature as the Hubble factor does not change. From the
definition of the
horizon, it is readily seen that%
\be%
\frac{dR_H} {dt} = H R_H-1. \ee%
In virtue of the above expression together with the Gibbs equation
one obtains

\be%
T \frac{d S}{dt} = -4\pi(\rho + P)R_H^2 , \ee%
whereby, the entropy transforms as%
\be%
S\rightarrow \widetilde{\overline{S}} = -[(1 + 3w)/(5 + 3w)]^2 S
\, . \label{stransf}
\ee
\\
Therefore, the entropy of the final non-phantom dark energy
dominated universe is proportional, but of the opposite the sign,
to the entropy of the original phantom universe. We might say that
the duality transformation ``quasi'' preserves the thermodynamics
of dark energy. This result is in keeping with the findings of
Section \ref{phcntw}.

%Before closing this paragraph, it is noteworthy that negative
%entropies are not so new after all; in principle monoatomic ideal
%gases should exhibit such a feature at sufficiently low
%temperatures \cite{appendixD}.

\section{Concluding remarks}
We have substantially extended the study of Pollock and Singh
\cite{pollock}, who considered the models of super--inflation
(characterized by $\dot{H} > 0$) of Starobinsky \cite{starobinsky}
and Shafi and Wetterich \cite{shafi}, by taking up models of
accelerated expansion (both with $\dot{H} > 0$ and with $\dot{H} <
0$) with $w$ constant as well as models with $w$ variable (such as
the model of Sami and Toporensky \cite{sami} and the Chaplygin gas
model \cite{chaplygin}).

We have found that, irrespective of whether $w$ is constant or
not, phantom fluids (associated with $\dot{H} > 0$) possess
negative entropy, transcend the holographic bound in the sense
that $\mid S\mid \geq S_{H}$ and their temperature and entropy
increase as the Universe expands. By contrast, non-phantom fluids
(associated with $\dot{H}<0$) have positive entropy, satisfy the
holographic bound, $S \leq S_{H}$, and their temperature and
entropy decrease with expansion. Except for the Chaplygin gas
model which violates the GSL at the earliest stage of dominating
the expansion, the GSL is fulfilled in all the cases.

It goes without saying that negative entropies are hard to
assimilate, they have no clear physical meaning in the context of
statistical mechanics; especially because the Einstein--Boltzmann
interpretation of entropy as a measure of the probability breaks
down. Systems of negative entropy appear to lie outside the
province of statistical mechanics as is currently formulated
\cite{populated,huang,mazenko}. However, if future cosmic
observations conclude that definitely $w <-1$, it will become
mandatory to generalize that subject accordingly. We wish to
emphasize that the laws of thermodynamics do not entail by
themselves that $S$ ought to be a positive quantity. The latter
will be found to be positive or negative only after combining
these laws with the equation (or equations) of state of the system
under consideration. What we believe to be at the core of
thermodynamics is the law that forbids the entropy of isolated
systems to decrease, this law encompasses the GSL for gravitating
systems with a horizon. Another interesting result is that, unlike
recent claims \cite{pedro}, the temperature of the phantom fluid
must be positive if the GSL is to be satisfied (see Eq.
(\ref{gibbs2})). In our view, this settles the dispute between
those rejecting phantom energy on grounds that it cannot be
physically acceptable because its entropy is negative
\cite{snegative} and those who evade negative entropies by
advocating negative temperatures instead \cite{pedro}. Both
approaches overlook the role of the event horizon.

As pointed out by Nojiri and Odintsov, it may well happen that in
the last stages of phantom dominated universes the scalar
curvature grows enough for quantum effects to play a
non-negligible role before the big rip. If so, the latter may be
evaded or at least softened \cite{nojiri}. While our study does
not incorporate such effects they should not essentially alter our
conclusions so long as the cosmic horizon persists.

It is noteworthy that the duality transformation $\rho + P
\rightarrow -(\rho + P)$ -see, e.g. \cite{duality}- along with
reversing the direction of time (i.e., $dt\rightarrow -dt$) leaves
Einstein's equations invariant and maps phantom cosmologies, with
both $H$ and $\dot{H}$ positive, into non-phantom cosmologies with
$H >0$ but $\dot{H} <0$ . However, this duality does not
necessarily extend to the thermodynamics of the respective
universes since future event horizons exist only when $\ddot{a} >
0$. Nevertheless, in the particular case of pole-like expansions,
Eq. (\ref{phexpansions}), the transformation preserves the
temperature while the entropy transforms according to Eq.
(\ref{stransf}) with the constraint $-5/3 < w <-1$. More general
instances will be considered elsewhere.

\acknowledgments{We are indebted to David Jou, Emili Elizalde,
Winfried Zimdahl and Orfeu Bertolami  for conversations on the
subject of this manuscript. G.I. acknowledges support from the
``Programa de Formaci\'{o} d'Investigadors de la UAB". This work
was partially supported by the Spanish Ministry of Science and
Technology under grant BFM2003-06033.}

\end{document}